\documentstyle[prl,aps,epsf]{revtex}         % 2 col mode

\tighten
\begin{document}
%
%\baselineskip=8.5mm  % preprint mode
%\draft               % preprint mode
% 2 col mode:
\twocolumn[\hsize\textwidth\columnwidth\hsize\csname@twocolumnfalse\endcsname
\title{Observation of vortex phase singularities in Bose-Einstein
condensates}

\author{S. Inouye, S. Gupta, T. Rosenband, A.P. Chikkatur,\\
A. G\"orlitz, T.L. Gustavson, A.E. Leanhardt,
D.E. Pritchard, and W. Ketterle }
\address{Department of Physics, Research Laboratory of
Electronics, and MIT-Harvard Center for Ultracold Atoms,\\
Massachusetts Institute of Technology, Cambridge, MA 02139, USA}

\date{\today}
\maketitle

\begin{abstract}

We have observed phase singularities due to vortex excitation in
Bose-Einstein condensates. Vortices were created by moving a
laser beam through a condensate. They were observed as
dislocations in the interference fringes formed by the stirred
condensate and a second unperturbed condensate. The velocity
dependence for vortex excitation and the time scale for
re-establishing a uniform phase across the condensate were
determined.
\end{abstract}


%03.75.Fi   Phase coherent atomic ensembles; quantum condensation phenomena
%03.75.Dg    Atom and neutron interferometry
%67.40.Vs   Vortices and turbulence

\vskip1pc
] %2 col mode
% NEXT TWO LINES NOT REVTEX!

%\narrowtext

Quantized vortices play a key role in the dynamics of superfluid
flow~\cite{Donn91}.  The nucleation of vortices determines the
critical velocity for the onset of dissipation at zero
temperature. In liquid helium, vortices are a source of friction
between the normal fluid and the superfluid. Multiple interacting
vortices can form a lattice or vortex tangle, depending on their
geometry and charge.

 Bose-Einstein condensates of dilute atomic gases offer a unique
 opportunity to study quantum hydrodynamics. The low density of the gas allows direct
comparison with first principle theories. A condensate is
characterized by a macroscopic wavefunction
$\psi(\vec{r})=\sqrt{\rho(\vec{r})}\exp(i\phi(\vec{r}))$, which
satisfies a non-linear Schr\"{o}dinger equation. The density
$\rho(\vec{r})$ and the velocity field $\vec{v}_{\rm s}(\vec{r})$
in the hydrodynamic equations can now be replaced by the square of
the wavefunction ($\rho(\vec{r})=|\psi(\vec{r})|^2$) and the
gradient of the {\it phase} of the wavefunction
\begin{equation}
    \vec{v}_{\rm s}(\vec{r})=\frac{\hbar}{m}\nabla \phi(\vec{r}) ,
    \label{eq:v-phi}
\end{equation}
where $m$ is the mass of the particle.

Recently, vortices in a Bose-Einstein condensate have been
realized experimentally and are currently under intensive
study~\cite{Matt99,Madi00,Ande00,Abos01}. In most of this work,
vortices were identified by observing the density depletion at
the cores. The velocity field was inferred only indirectly, with
the exception of the work on circulation in a two-component
condensate~\cite{Matt99}. The flow field of a vortex can be
directly observed when the phase of the macroscopic wavefunction
is measured using interferometric techniques.  In this work, we
created one or several vortices in one condensate and imaged its
phase by interfering it with a second unperturbed condensate
which served as a local oscillator.

Interferometric techniques have previously been applied either to
simple geometries such as trapped or freely expanding
condensates~\cite{Andr97,Sims00,Bloc00}, or to read out a phase
imprinted by rf- or optical fields~\cite{Matt99,Hall98,Dens00}.
Here we apply an interferometric technique to visualize turbulent
flow.

\begin{figure}[htbf]
\epsfxsize=70mm \centerline{\epsfbox{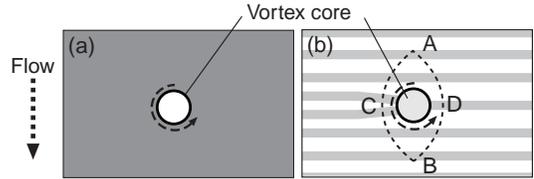}}\vspace{0.5cm}
\caption{ Density (a) and phase (b) profile of a moving
condensate with singly-charged ($n=1$) vortex. The density
profile shows the vortex core, whereas the phase pattern features
a fork-like dislocation at the position of the vortex.
Interference between two initially separated, freely expanding
condensates produces exactly the same pattern as shown in (b), if
one of the condensate contains a vortex.} \label{fig:singularity}
\end{figure}

The line integral of Eq.~(\ref{eq:v-phi}) around a closed path
gives the quantization of circulation:
\begin{equation}
\int \vec{v}(\vec{r}) \cdot d\vec{r} = \frac{\hbar}{m}
\left(\phi(\vec{r}_{\rm f}) - \phi(\vec{r}_{\rm i})\right) .
\end{equation}
If the path is singly connected, there is no circulation. If the
path is multiply connected (like around a vortex core) the
circulation can take values $nh/m$ (integer multiples of $ h /
m$), since the phase is only defined modulo $2\pi$. As a result,
the phase accumulated between two points A and B can be different
depending on the path (Fig.~\ref{fig:singularity}). The integer
quantum number $n$ is called the charge of the vortex.  When the
phase gradient is integrated along a path to the left of the
vortex (path ACB), the accumulated phase differs by $2n\pi$ from
the path to the right (ADB).

This phase difference can be visualized with interferometric
techniques. When two condensates with relative velocity $v$
overlap, the total density shows straight interference fringes
with a periodicity $h/mv$. If one of the condensates contains a
vortex of charge $n$, there are $n$ more fringes on one side of
the singularity than on the other side
(Fig.~\ref{fig:singularity}b). The change in the fringe spacing
reflects the velocity field of the vortex. An observation of this
fork-like dislocation in the interference fringes is a clear
signature of a vortex~\cite{Bold98}.

Our setup for the interferometric observation of vortices is
essentially a combination of two experiments conducted in our lab
in the past \cite{Andr97,Rama99}. Briefly, laser cooled sodium
atoms were loaded into a double-well potential and further cooled
by rf-induced evaporation below the BEC transition temperature.
The double-well potential was created by adding a potential hill
at the center of a cigar-shaped magnetic trap. For this,
blue-detuned far off-resonant laser light ($532\,{\rm nm}$) was
focused to form an elliptical $75\,\mu{\rm m} \times 12\,\mu{\rm
m}$ (FWHM) light sheet and was aligned to the center of the
magnetic trap with the long axis of the sheet perpendicular to the
long axis of the condensate. The condensates produced in each
well were typically $20\,\mu{\rm m}$ in diameter and
$100\,\mu{\rm m}$ in length. The height of the optical potential
was $\sim 3\,{\rm kHz}$, which was slightly larger than the
chemical potential of the condensate. A more intense light sheet
would have increased the distance between the condensates, thus
reduced the fringe spacing~\cite{Andr97}.

\begin{figure}[htbf]
\epsfxsize=70mm \centerline{\epsfbox{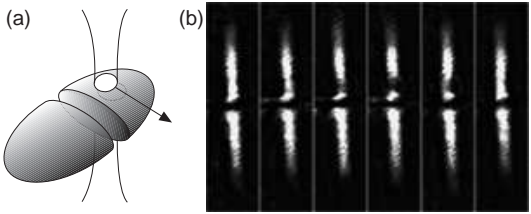}}\vspace{0.5cm}
\caption{Schematic (a) and phase-contrast images (b) of the
condensates used for the experiment. A blue-detuned laser beam
(not shown in the figure) was focused into a light sheet
separating the two condensates in the magnetic trap. Another
tightly focused laser beam was swept through one of the
condensates (the upper one in image (b)) to excite vortices. The
intensity of each laser beam was a factor of four higher than in
the experiments to enhance the depleted regions in the images.
The images in (b) have a field of view of  $100\,\mu{\rm m} \times
380\,\mu{\rm m}$. For each image, the stirrer was advanced from
left to right by $5\,\mu{\rm m}$.} \label{fig:slashedBEC}
\end{figure}

After two condensates each containing $\sim 1 \times 10^6$ atoms
in the $F=1, m_F=-1$ state were formed in the double-well
potential, we swept a second blue-detuned laser beam through one
of the condensates using an acousto-optical deflector
(Fig.~\ref{fig:slashedBEC}). The focal size of the sweeping laser
beam ($12\,\mu{\rm m} \times 12\,\mu{\rm m}$, FWHM) was close to
the width of the condensate. The alignment of this beam was
therefore done using an expanded condensate in a weaker trap
where the beam profile created a circular ``hole'' in the
condensate density distribution.

After sweeping the beam once across the ``sample'' condensate,
the magnetic and optical fields were switched off and the two
condensates expanded and overlapped during $41\,{\rm ms}$
time-of-flight. The atoms were then optically pumped into the $F =
2$ hyperfine ground state for $80\,\mu s$ and subsequently probed
for $20\,\mu$s by absorption imaging tuned to the $F = 2$ to
$F^{\prime} = 3$ cycling transition.

Obtaining high contrast interference fringes ($\sim 70\% $) across
the whole cloud required attention to several key factors. First,
standard absorption imaging integrates along the line of sight,
which was vertical in our experiment. Any bending or distortions
of the interference fringes along the direction of observation
would result in a loss of contrast. This was avoided by
restricting the absorption of the probe light to a thin
horizontal slice. The optical pumping beam was focused into a
light sheet of adjustable thickness (typically $100\,\mu{\rm m}$,
which is about 10\% of the diameter of the cloud after the
time-of-flight) and a width of a few millimeters. This pumping
beam propagated perpendicularly to the probe light and parallel to
the long axis of the trap. Second, the number of atoms in the
condensates had to be reduced to about $1\times 10^6$
(corresponding to a chemical potential $\mu \sim 2.5\,{\rm
kHz}$). Higher numbers of atoms resulted in a severe loss of
contrast, even if we detuned the probe beam to reduce optical
density. We suspect that at high density, the two condensates do
not simply interpenetrate and interfere, but interact and
collide.  Third, high spatial homogeneity of the probe beam was
important to obtain absorption images with low technical noise.
In some of our experiments, the probe beam position was actively
scanned to smooth the beam profile. Fourth, the intensity of the
sweeping blue-detuned beam was adjusted so that the height of the
optical potential was a fraction (typically one half) of the
chemical potential of the condensate. Higher intensity of the
sweeping beam resulted in reduced interference fringe contrast,
probably due to other forms of excitations.

\begin{figure}[htbf]
\epsfxsize=95mm \centerline{\epsfbox{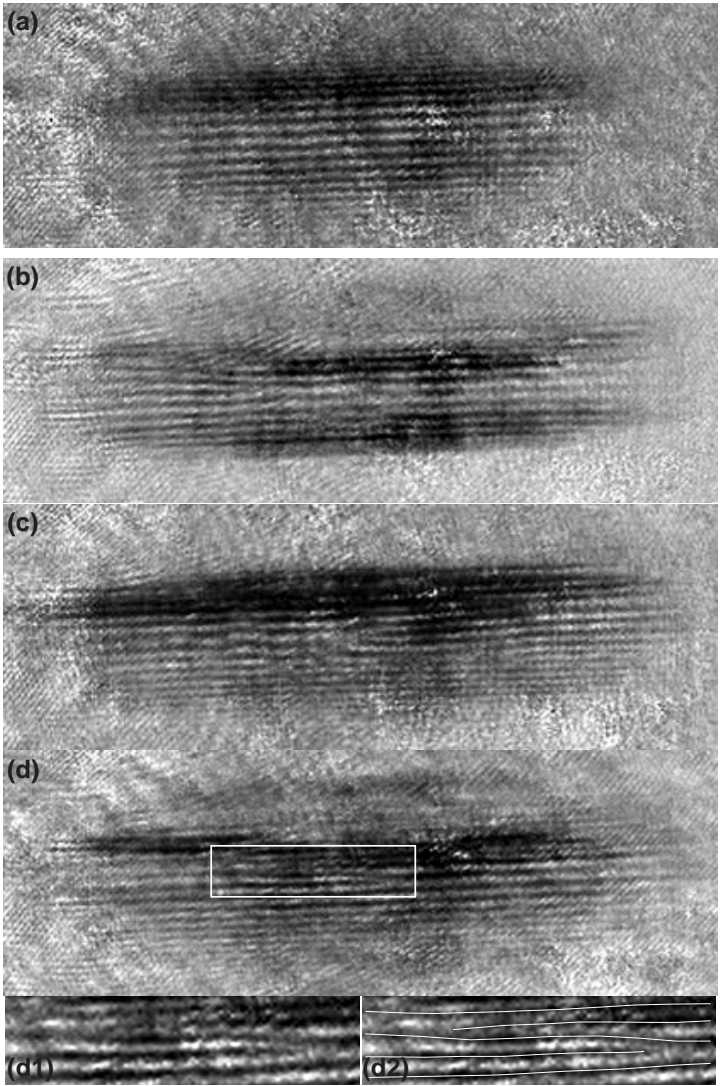}} \vspace{0.5cm}
\caption{ Observation of the phase singularities of vortices
created by sweeping a laser beam through a condensate. Without
the sweep, straight fringes of $\sim 20\,\mu{\rm m}$ spacings were
observed (a), while after the sweep, fork-like dislocations
appeared (b-d). The speed of the sweep was $1.1\,\mu{\rm m/ms}$,
corresponding to a Mach number of $\sim 0.18$. The field of view
of each image is $1.1\,{\rm mm} \times 0.38\,{\rm mm}$. Fig.\ (d)
shows a pair of dislocations with opposite circulation
characteristic of a vortex pair.  At the bottom, magnified images
of the fork-like structures are shown (d1) with lines to guide
the eye (d2). The orientation of the condensates is the same as
in Fig.~\ref{fig:slashedBEC}(b).} \label{fig:forks}
\end{figure}

Images of interfering condensates show a qualitative difference
between stirred (Fig.~\ref{fig:forks}(b-d)) and unperturbed
states (Fig.~\ref{fig:forks}(a)). Fork-like structures in the
fringes were often observed for stirred condensates, whereas
unperturbed condensates always showed straight fringes.  The
charge of the vortices can be determined from the fork-like
pattern. In Fig.\ 3(b), vortices were excited in the condensate
on top, and the higher number of fringes on the left side
indicates higher relative velocity on this side, corresponding to
counterclockwise flow. Fig.\ 3(c) shows a vortex of opposite
charge. The double fork observed in Fig.\ 3(d) represents the
phase pattern of a vortex pair. Multiply charged vortices, which
are unstable against the break-up into singly charged vortices,
were not observed.

Theoretical studies of the superfluid flow around moving objects
predict dissipationless flow below a critical
velocity~\cite{Donn91}. Above this velocity, vortices of opposite
circulation are created on the two sides of the moving object and
give rise to a drag force~\cite{Fris92}. A recent experiment in
our group found the onset of dissipation at a critical Mach
number of $v_c/c_s \sim 0.1$~\cite{Onof00}. Dissipation at low
velocities can not only occur by vortex shedding, but also by the
creation of phonons in the low density regions of the
condensate~\cite{Fedi00}.  The direct observation of vortices at
similar Mach numbers (Fig.\ 3) provides strong evidence that
vortices play a major role in the onset of dissipation at the
critical velocity.

\begin{figure}[htbf]
\epsfxsize=65mm \centerline{\epsfbox{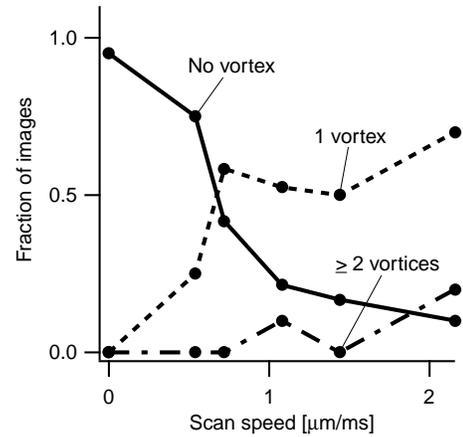}}\vspace{0.5cm}
\caption{ Velocity dependence of vortex excitation. The fraction
of images with zero (solid line), one (dashed line), and two or
more vortices (dash-dotted line) are plotted versus the speed of
the sweep. After the sweep, the atoms were released from the trap
without delay. The total number of evaluated images was 50.
Ambiguous low contrast images were excluded; therefore, the sum of
the fractions is less than one.}\label{fig:scanspeed}
\end{figure}

By varying the speed of the laser beam sweep, we determined the
velocity dependence of the vortex nucleation process. Due to the
turbulent nature of the flow, every image was different even if
they were taken under the same experimental conditions. Thus the
images were classified by counting the number of vortices and the
fractions were plotted versus the speed of the sweep (Fig.\
\ref{fig:scanspeed}). The classification was done after putting
images in random order to eliminate a possible ``psychological
bias.'' The plot suggests that the nucleation of vortices requires
a velocity of $\sim 0.5\,\mu{\rm m/ms}$, corresponding to a Mach
number $v_c/c_s \sim 0.08$, consistent with our previous
measurement~\cite{Onof00}. However, a direct comparison is not
possible due to different geometries---in the present experiment,
the stirrer was swept along the radial direction of the
condensate and almost cut the cloud completely, whereas in the
previous experiment, the stirrer moved along the axial direction
of an expanded condensate.

\begin{figure}[htbf]
\epsfxsize=65mm \centerline{\epsfbox{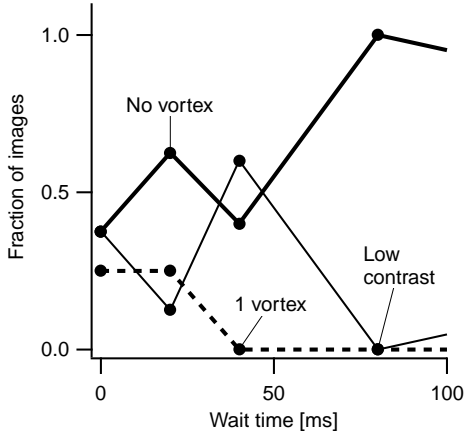}}\vspace{0.5cm}
\caption{ Relaxation of a condensate towards uniform phase.  The
fraction of images with zero (thick solid line) and one (dashed
line) vortex and with low contrast (thin solid line) are plotted
versus the waiting time after the laser beam sweep ($v/c_{\rm
s}\sim 0.09$). The total number of images used for creating this
plot was 33.}\label{fig:waittime}
\end{figure}

Previous experiments have dramatically demonstrated the
robustness of the long-range coherence of the
condensate~\cite{Andr97,Hall98}. The interferometric technique
used here is a sensitive way to assess whether a condensate has
the assumed ground state wave function which is characterized by a
uniform phase. Sweeping through the condensate excites turbulent
flow. By delaying the release of the atoms from the trap by a
variable amount of time, we can study the relaxation of the
condensate towards its ground state. Fig.~\ref{fig:waittime}
shows that the condensate completely recovers its uniform phase
after $50-100\,{\rm ms}$. Vortices have disappeared after $\sim
30\,{\rm ms}$. Of course, these measurements depend crucially on
the specific geometry of the cloud, but they do indicate typical
time scales. The sensitivity of this method was illustrated by the
following observation:  in a weaker trap, we saw an oscillation in
time between images with straight high contrast fringes and images
with low contrast fringes. This was probably due to the
excitation of a sloshing motion along the weak axis of the
condensate.

For interferometric detection of vortices, two different
techniques have been discussed. The one employed here uses a
separate condensate as a local oscillator.  The other alternative
is to split, shift and recombine a single condensate with
vortices. In this case, the interference pattern is more
complicated because all singularities and distortions appear
twice.  The simulations in Ref.\ \cite{Dobr99} show that the
self-interference technique produces more complicated fringe
patterns. After completion of this work, we learned that this
second technique was used in ENS, Paris to observe the phase
pattern of a single vortex~\cite{DaliTalk}.

In conclusion, we have studied vortex excitation in Bose-Einstein
condensates using an interferometric technique.  This technique is
suited for the study of complicated superfluid flows, e.g., when
multiple vortices with opposite charges are present.  We have
obtained a clear visualization of vortices as topological
singularities, confirmed the role of vortices in the onset of
dissipation near the critical velocity, and observed the
relaxation of a stirred condensate towards a state with uniform
phase.

The field of Bose-Einstein condensation combines atomic and
condensed matter physics.  This aspect is illustrated by this
work where an atomic physics technique, matter wave
interferometry, was used to probe the nucleation of vortices, a
problem of many-body physics.  There are many issues of vortex
physics which remain unexplored, including vortices in
two-dimensional condensates (condensates in lower dimensions were
recently realized in our laboratory~\cite{Gorl01}), pinning of
vortices by additional laser beams, and interactions between
vortices.

This work was supported by NSF, the ONR, ARO, NASA, and the David
and Lucile Packard Foundation. A.E.L.\ and A.P.C.\ acknowledge
additional support by fellowships from NSF, and JSEP,
respectively.

\end{document}